\documentclass[prb,twocolumn,aps,showpacs,fixfloats]{revtex4}
\usepackage{graphicx}
\usepackage{bm}
\usepackage{amsmath,amssymb}
\usepackage{subfigure}
\usepackage{float}
\usepackage{latexsym}
\usepackage{color}
\usepackage{enumerate}
\usepackage{pdfpages}
\usepackage{tikz}
\usepackage{hyperref}
\usepackage{relsize}

\begin{document}
\newcommand{\s}{\scriptscriptstyle}
\newcommand{\uu}{\uparrow \uparrow}
\newcommand{\ud}{\uparrow \downarrow}
\newcommand{\du}{\downarrow \uparrow}
\newcommand{\dd}{\downarrow \downarrow}
\newcommand{\ket}[1] { \left|{#1}\right> }
\newcommand{\bra}[1] { \left<{#1}\right| }
\newcommand{\bracket}[2] {\left< \left. {#1} \right| {#2} \right>}
\newcommand{\vc}[1] {\ensuremath {\bm {#1}}}
\newcommand{\tr}{\text{Tr}}
\newcommand{\Trans}{\ensuremath \Upsilon}
\newcommand{\Refl}{\ensuremath \mathcal{R}}

\title{Suppression of the Landau-Zener transition probability by a weak classical  noise }

\author{Rajesh K. Malla, E. G. Mishchenko,  and M. E. Raikh}

\affiliation{ Department of Physics and
Astronomy, University of Utah, Salt Lake City, UT 84112}

\begin{abstract}
When the drive which causes the level crossing in a qubit is slow,
the probability, $P_{\s LZ}$, of the Landau-Zener transition is close to $1$.
We show that in this regime,
which is most promising for applications,
the noise due to the coupling to the environment,
reduces the average $P_{\s LZ}$.
At the same time, the survival probability, $1-P_{\s LZ}$, which is exponentially small for a slow drive, can be completely dominated by noise-induced correction.
Our main message  is that the effect of a weak classical noise can be captured
analytically by treating it as a perturbation in the Schr{\"o}dinger equation.
This allows us to study the dependence of the noise-induced correction to $P_{\s LZ}$ on the correlation time of the noise. As this correlation time
exceeds the
bare Landau-Zener transition time, the effect of noise becomes negligible.  We 
consider two conventional realizations of
noise: gaussian noise and telegraph noise.

\end{abstract}

\maketitle

\section{Introduction}

Theoretical papers on coherent manipulation of the
quantum states of a qubit can be divided into two groups.
At the focus of the first group, see e.g. Refs. \onlinecite{Lim1991+++,Garanin2002+++,Berry2009+++,Bason2012+++,delCampo2013+++,Zhang2013+++,Ban2014+++,Polleti2016+++,Funo2017+++},
is a quest for  ``superadiabaticity",
which is an optimal protocol of drive-induced crossing of the energy levels.
Following this protocol, at the end of the evolution,
the final state of a qubit is as close as possible to the adiabatic
ground state. If the time variation of the energy levels is linear, $\pm vt/2$,
where $v$ is the drive velocity,
the degree of adiabaticity is given by the celebrated Landau-Zener (LZ)
formula\cite{Landau1932,Zener1932}
\begin{equation}
\label{PLZ1}
P_{\s LZ}=1-Q_{\s LZ},  ~~~~Q_{\s LZ}= \exp{\left\{-\frac{2\pi J^2}{v}\right\}},
\end{equation}
where $J$ is the tunnel splitting of the levels at the crossing point.
The meaning of $P_{\s LZ}$ is the probability to find the system,
which is in  $\uparrow$ state at $t\rightarrow -\infty$,
in the state $\downarrow$ at $t\rightarrow \infty$.
Correspondingly, the meaning of $Q_{\s LZ}$ is the ``survival" probability
to find the system in the initial state.

The value $P_{\s LZ}$ serves as an estimate of the  degree
of adiabaticity achievable when a two-level system
is forced through an avoided crossing.
In this regard, ``superadiabatic" protocol minimizes
the survival probability.

In the papers of the second group, see e.g. Refs.
[\onlinecite{Galperin2008+,Nalbach2009+,Pekola2011+,Ziman2011+,Vavilov2014+,Nalbach2014+,Nalbach2015+,Ogawa2017+}],
 the drive is assumed to be strictly linear.
The subject of study is the effect of coupling
of the qubit levels to the environment on the probability
of the Landau-Zener transition.

A common approach to the study of the effect of environment (thermal bath)  on
the LZ transition is to add to the Hamiltonian of the two-level system
the Hamiltonian of the bath and the Hamiltonian of the linear coupling
of the bath to the two-level system. After that, the equations of motion
for the density matrix are cast in the form of master equations.
This is achieved by generalizing the Lindblad approach of Bloch-Redfield
approach developed for stationary two-level systems to the case of
time-dependent Hamiltonian.
The resulting closed system of master equations is solved numerically.\cite{Galperin2008+,Nalbach2009+,Pekola2011+,Ziman2011+,Vavilov2014+,Nalbach2014+,Nalbach2015+,Ogawa2017+}
This numerics sometimes reveals a peculiar dependence\cite{Ziman2011+} of the dynamics of the
LZ transition  on the noise frequency and intensity or, more precisely, on temperature.

The message of the present paper is that the effect of a weak classical noise
can be studied analytically by treating it as
perturbation in the Schr{\"o}dinger equation. This allows to study
the dependence of the noise-induced correction to $P_{\s LZ}$ on
the correlation time of the noise. The situation when this correction plays
a crucial role is strong-coupling limit, $J\gg v^{1/2}$, when the bare LZ
transition probability is exponentially close to $1$.
In this limit, the bare survival probability, $Q_{\s LZ}$,
is exponentially small. We will show that the correction to $P_{\s LZ}$
is negative and does not contain the exponential factor $\exp{\left[-(2\pi J^2)/v\right]}$.
Thus, even a weak noise can dominate $Q_{\s LZ}$. We analyze the noise-induced correction for
the two realizations of the noise: gaussian noise and the telegraph noise.
\begin{figure}
\includegraphics[scale=0.13]{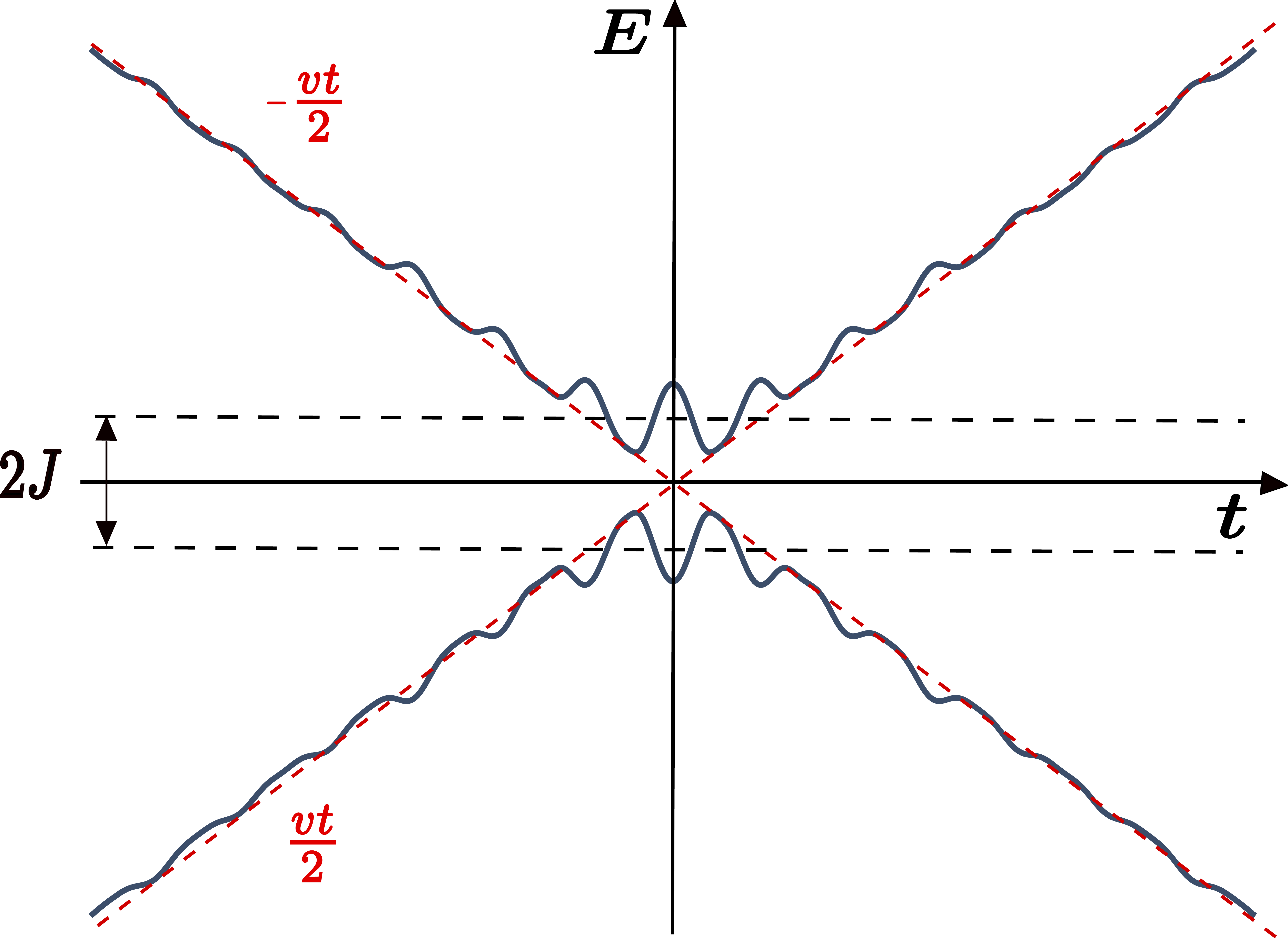}
\caption{(Color online) Schematic illustration of the Landau-Zener transition in the presence
of a weak transverse noise. The noise causes random fluctuations of the gap, $2J$. When the gap
is much bigger than $v^{1/2}$, where $v$  is the sweep velocity, the LZ transition is almost
fully adiabatic, so that the ``survival" probability, $Q_{\s LZ}$, to stay on the initial diabatic
level is exponentially small. Then, even a weak noise yields a dominant contribution to $Q_{\s LZ}$.}
\label{figure1}
\end{figure}
\section{Perturbative solution of the Schr{\"o}dinger equation in the presence of noise}
Denote with $a_{\s \uparrow}$, $a_{\s \downarrow}$ the
amplitudes to find a driven system in the $\uparrow$ and $\downarrow$ states, respectively. In the presence of random
$\delta J(t)$, modeling the noise, these amplitudes satisfy the
following system of equations
\begin{equation}
\label{system}
\begin{cases}
i{\dot a}_{\s \uparrow}=\frac{vt}{2}a_{\s \uparrow}+[J+\delta J(t)]a_{\s \downarrow},  \\
i{\dot a}_{\s \downarrow}=-\frac{vt}{2}a_{\s \downarrow}+[J+\delta J(t)]a_{\s \uparrow}. \\
\end{cases}
\end{equation}
In the absence of noise, two linearly independent solutions of the system Eq. (\ref{system}) have the form
\begin{equation}
\label{solution11}
\begin{cases}
a_{\s \uparrow}^{\s (1)}=D_\nu(z),\\
a_{\s \downarrow}^{\s (1)}=-i\sqrt{\nu} D_{\nu-1}(z),
\end{cases}
\end{equation}
\begin{equation}
\label{solution12}
\begin{cases}
a_{\s \uparrow}^{\s (2)}=D_\nu(-z),\\
a_{\s \downarrow}^{\s (2)}=-i\sqrt{\nu} D_{\nu-1}(-z),
\end{cases}
\end{equation}
where $D_\nu(z)$ is the parabolic cylinder function\cite{Bateman1955} of the argument, $z$,  defined as $z=\sqrt{v}e^{i\pi/4}t$, while
the index $\nu$ is given by
\begin{equation}
\label{nu1}
\nu=-\frac{iJ^2}{v}.
\end{equation}
The solution Eq. \ref{solution11} satisfies the ``right" initial condition $a_{\downarrow}^{\s (1)}(-\infty)=0$, i.e. that the system
is initially in the state~$\uparrow$.

In the presence of noise, we search for the corrections to the amplitudes, $a_{\s \uparrow}^{\s (1)}$ and $a_{\s \downarrow}^{\s (1)}$,
in the form of the linear combination
\begin{equation}
\label{Matrix1}
\begin{pmatrix}
\delta a_{\s \uparrow} \\ \delta a_{\s \downarrow}
\end{pmatrix} = c_{\s 1}(t)\begin{pmatrix}
 a_{\s \uparrow}^{\s (1)} \\  a_{\s \downarrow}^{\s (1)}
\end{pmatrix}+c_{\s 2}(t)\begin{pmatrix}
 a_{\s \uparrow}^{\s (2)} \\  a_{\s \downarrow}^{\s (2)}
\end{pmatrix}.
\end{equation}
Substituting this form into  Eq. (\ref{system}) and keeping only
$a_{\s \uparrow}^{\s (1)}$, $a_{\s \downarrow}^{\s (1)}$ in the terms
proportional to $\delta J$ we arrive to the following linear system of equations for ${\dot c_{\s 1}(t)}$ and ${\dot c_{\s 2}(t)}$
\begin{equation}
\label{systemmodified}
\begin{cases}
i\left({\dot c_{\s 1}(t)} a_{\s \uparrow}^{\s (1)}+{\dot c_{\s 2}(t)} a_{\s \uparrow}^{\s (2)} \right)=\delta J(t)a_{\s \downarrow}^{\s (1)},\\
i\left({\dot c_{\s 1}(t)} a_{\s \downarrow}^{\s (1)}+{\dot c_{\s 2}(t)} a_{\s \downarrow}^{\s (2)} \right)=\delta J(t)a_{\s \uparrow}^{\s (1)}.
\end{cases}
\end{equation}
Taking into account the initial conditions $c_{\s 1}(-\infty)=0$ and
$c_{\s 2}(-\infty)=0$, we find the expressions for $c_{\s 1}$ and
$c_{\s 2}$
\begin{eqnarray}
&\hspace{-3mm}c_{\s 1}(t)=-i\hspace{-1mm}\int\limits_{-\infty}^{t}\hspace{-1mm} dt' \delta J(t')\frac{a_{\s \downarrow}^{\s (1)}(t')a_{\s \downarrow}^{\s (2)}(t')-a_{\s \uparrow}^{\s (1)}(t')a_{\s \uparrow}^{\s (2)}(t')}{a_{\s \uparrow}^{\s (1)}a_{\s \downarrow}^{\s (2)}-a_{\s \uparrow}^{\s (2)}a_{\s \downarrow}^{\s (1)}}, \label{Camplitude1}\\
&\hspace{-3mm}c_{\s 2}(t)=-i\int\limits_{-\infty}^{t} dt' \delta J(t')\frac{[a_{\s \downarrow}^{\s (1)}(t')]^2-[a_{\s \uparrow}^{\s (1)}(t')]^2}{a_{\s \uparrow}^{\s (1)}a_{\s \downarrow}^{\s (2)}-a_{\s \uparrow}^{\s (2)}a_{\s \downarrow}^{\s (1)}}. \label{Camplitude2}
\end{eqnarray}
It is easy to see that the denominator in Eqs. (\ref{Camplitude1}), (\ref{Camplitude2}) is a time independent constant. This is the consequence of the relation
\begin{equation}
\label{Wronskianrelation}
J \left(a_{\s \uparrow}^{\s (1)}a_{\s \downarrow}^{\s (2)}-a_{\s \uparrow}^{\s (2)}a_{\s \downarrow}^{\s (1)} \right)=i \left( {\dot a_{\s \uparrow}^{\s (1)}}a_{\s \uparrow}^{\s (2)}-{\dot a_{\s \uparrow}^{\s (2)}}a_{\s \uparrow}^{\s (1)}\right),
\end{equation}
which straightforwardly follows from the system Eq. (\ref{system}).
The expression in the right-hand side is a Wronskian, the value of
which is known\cite{Bateman1955}
\begin{equation}
\label{Wronskianvalue}
D_\nu(z)\frac{d}{dz}D_\nu(-z) - D_\nu(-z)\frac{d}{dz}D_\nu(z) = \frac{(2\pi)^{1/2}}{\Gamma(-\nu)}.
\end{equation}
Here $\Gamma(-\nu)$ is the Gamma-function.

The exact expression for the survival probability
is ${Q}_{\s LZ}=|a_{\s \uparrow}(\infty)/a_{\s \uparrow}(-\infty)|^2$. Using Eq. (\ref{Matrix1}), we can express this probability, with noise
taken into account to the lowest order, via the bare survival probability as follows

\begin{multline}
\label{QLZ1}
{Q}_{\s LZ}=|1+c_1(\infty)|^2 e^{-2\pi|\nu|}\\
+2\text{Re}\big[(1+c_1(\infty))^*c_2(\infty)\big]e^{-\pi|\nu|}+|c_2(\infty)|^2.
\end{multline}
The latter expression illustrates our main point, namely, when the
bare survival probability is exponentially small, the net survival probability is dominated by the noise-induced correction, $|c_2(\infty)|^2$.
The analytical expression for this correction follows from Eq. (\ref{Camplitude2}). It should be averaged over the noise realizations. This averaging is carried out in the next Section.

\section{Averaging over the noise realizations}

The strength and the correlation time of the noise are encoded in the correlator
defined as
\begin{equation}
\label{timecorrelator}
\langle \delta J(t_1)\delta J(t_2)\rangle =(\delta J)^2 K(t_1-t_2),
\end{equation}
where $\delta J$ is the r.m.s.  noise magnitude and $K(0)=1$.

Using Eqs. (\ref{Wronskianrelation}), and (\ref{Wronskianvalue}), the average survival probability, $\langle{Q}_{\s LZ}\rangle=\langle\vert c_2(\infty)\vert^2\rangle $, can be expressed via the correlator as follows
\widetext
\begin{equation}
\label{bigeq1}
\langle|c_2(\infty)|^2\rangle = \frac{(\delta J)^2}{2\sinh \pi |\nu|} \int\limits_{-\infty}^{\infty}dt_1\int\limits_{-\infty}^{\infty} dt_2 K(t_1 -t_2)\Bigg\{ \Big[a_{\s \downarrow}^{\s (1)}(t_1)\Big]^2-\Big[a_{\s \uparrow}^{\s (1)}(t_1)\Big]^2\Bigg\}\Bigg\{ \Big[a_{\s \downarrow}^{\s (1)}(t_2)\Big]^2-\Big[a_{\s \uparrow}^{\s (1)}(t_2)\Big]^2\Bigg\}^*,
\end{equation}
\endwidetext
where we used the identity $|\Gamma (-\nu)|^2=\pi/|\nu|\sinh(\pi|\nu|)$.

To evaluate the double integral, we take advantage of the fact that, without noise,
the transition probability is close to $1$, which implies that
the parameter $|\nu|$ is big, $|\nu|\gg 1$.
This, in turn, justifies using the semiclassical asymptotes for
the parabolic cylinder functions not only for big, but, in fact,
for {\em all} values of the argument. The asymptotic forms of
the parabolic cylinder functions valid at large $|\nu|$ and arbitrary $t$ can be found in Ref.
\onlinecite{Luo2017}. Using these asymptotes, for the  combination $\Big[a_{\s \downarrow}^{\s (1)}(t)\Big]^2-\Big[a_{\s \uparrow}^{\s (1)}(t)\Big]^2$ which enters into Eq. (\ref{bigeq1}), one obtains
\begin{multline}
\label{numeratorbigeq}
\Big[a_{\s \downarrow}^{\s (1)}(t)\Big]^2-\Big[a_{\s \uparrow}^{\s (1)}(t)\Big]^2=D_{\nu}^2\Big(\sqrt{v}e^{i\frac{\pi}{4}}t\Big)+\nu D_{\nu-1}^2\Big(\sqrt{v}e^{i\frac{\pi}{4}}t\Big)\\
\approx \frac{\frac{vt}{2}\exp\Big(\frac{\pi |\nu|}{2}\Big)}{\Big(J^2+ \frac{v^2t^2}{4}\Big)^{1/2}} \exp[-2i\Phi(t)],
\end{multline}
where $\Phi(t)$ is the semiclassical phase
\begin{equation}
\label{semiclassicalphase}
\Phi(t)=\int\limits_0^t dt' \Big[J^2+\frac{v^2t'^2}{4}\Big]^{1/2}.
\end{equation}
Due to $|\nu|$ being large, the term corresponding to $\exp(2i\Phi(t))$
in  Eq. (\ref{numeratorbigeq}) is exponentially suppressed. The denominator
in the prefactor is conventional for  semiclassics. Appearance of $t$ in the numerator
can be simply illustrated by substituting $a_{\s \downarrow}^{\s (1)}(t)\propto
\exp(-i\Phi(t))$ into the system Eq. (\ref{system}). This will yield the relation

\begin{equation}
\label{relation}
\Big[a_{\s \downarrow}^{\s (1)}(t)\Big]^2-\Big[a_{\s \uparrow}^{\s (1)}(t)\Big]^2
\approx -\frac{vt}{2J}\Big[a_{\s \uparrow}^{\s (1)}(t)\Big]^2.
\end{equation}
For the further evaluation of the
double integral in Eq. (\ref{bigeq1}), it is convenient to switch from time domain to the frequency domain,
as it is illustrated in the next section.
\begin{figure}
\includegraphics[scale=0.2]{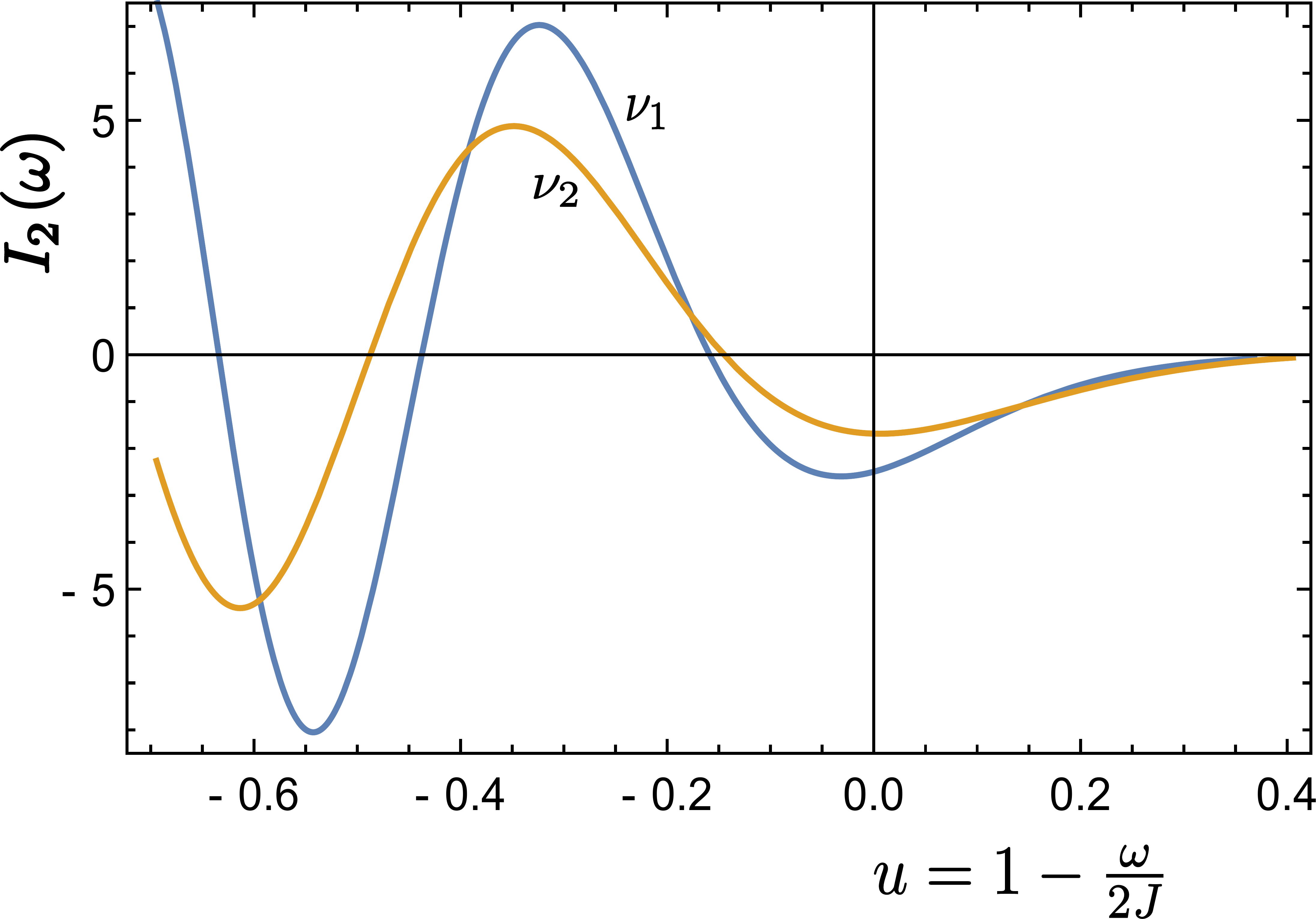}
\caption{(Color online) The integral $I_2(\omega)$ (in the units $4i(2\pi)^{1/2}/J^3$) is plotted from
Eq. (\ref{smallomegaintegral}) versus the dimensionless variable, $u=\left(1-\omega/2J\right)$, for two values of the dimensionless parameter $\nu$:
$\nu=3$ (orange) and $\nu=3.5$ (blue). For negative $u$, $I(\omega)$ oscillates and reproduces
the semiclassical result Eq. (\ref{bigomegaIntegral}) after the first maximum.   For positive $u$ it falls off exponentially. Despite the $u>0$-tail is slim,
it is responsible the survival probability when the noise is slow. }
\label{Airyplot}
\end{figure}
\section{Calculation of $Q_{\s LZ}$ in the frequency domain}
Denote with ${\tilde K}$ the Fourier transform of the correlator Eq. (\ref {timecorrelator})
\begin{equation}
\label{frequencycorrelator}
K(t)=\int\limits_{-\infty}^{\infty} \tilde{K}(\omega) e^{i\omega t}d\omega.
\end{equation}
Upon substituting Eq. (\ref {frequencycorrelator}) into Eq. (\ref {bigeq1}), the integrations
over $t_1$ and $t_2$ get decoupled and we obtain
\begin{equation}
\label{Fourierprobability}
\langle|c_2(\infty)|^2\rangle=\frac{e^{\pi|\nu|}}{2\sinh \pi |\nu|} (\delta J)^2\int\limits_{-\infty}^{\infty} d\omega \tilde{K}(\omega) |I(\omega)|^2,
\end{equation}
where $I(\omega)$ is given by
\begin{equation}
\label{FourierIntegral}
I(\omega)=\int\limits_{-\infty}^{\infty} dt~ \frac{\frac{vt}{2}}{\Big(J^2+ \frac{v^2t^2}{4}\Big)^{1/2}}\exp\Big[ i\Big(\omega t -2\Phi(t) \Big)\Big].
\end{equation}
Analytical form of $I(\omega)$ depends on the frequency domain. For
high $\omega$ one can use the steepest descent method.
The exponent in Eq. (\ref{FourierIntegral}) has two extrema at $t=\pm t_{\omega} $,
where
\begin{equation}
\label{textrema}
t_{\omega}=\frac{2}{v}\left[\frac{\omega^2}{4}-J^2\right]^{1/2}.
\end{equation}
Expanding the exponent near these extrema and taking into account that
$\partial^2 \Phi/\partial t^2=v^2t/4(J^2 +v^2t^2/4)^{1/2}$, after combining the
two contributions, we obtain
\begin{equation}
\label{bigomegaIntegral}
I(\omega)\Big|_{\s \omega > 2J}=I_1(\omega)=i\frac{2^{3/2}\pi^{1/2}t_{\omega}^{1/2}}{|\omega|^{1/2}}\sin\Big[\omega t_{\omega} -2\Phi(t_{\omega})+\frac{\pi}{4} \Big].
\end{equation}
The above result applies when the argument of sine is big. For $|\nu|\gg 1$
this requirement is already satisfied when $\omega$ exceeds $2J$ only slightly.
Indeed,
the criterion $\omega t_{\omega}\gg 1$ can be cast in the form
\begin{equation}
\label{criterion}
\left(\omega -2J\right) \gg \frac{J}{|\nu|^2}.
\end{equation}
Physically, this criterion means that the direct absorption (emission) of a
noise ``quantum", say, a phonon, if the noise is due to lattice vibrations, is allowed.

For frequencies $\omega <2J$ the behavior of $I(\omega)$ exhibits a sharp
cutoff as the difference $2J-\omega$ grows. It appears that, in order
 to capture this cutoff, it is sufficient to replace $\Phi (t)$ by its
 small-$t$ expansion, namely
 \begin{equation}
 \label{expansion}
 \Phi(t)\approx Jt +\frac{v^2t^3}{12J}.
 \end{equation}
 One can also neglect $v^2t^2/4$ in the denominator of
 Eq. (\ref{FourierIntegral} ). After that, $I(\omega)$  reduces to the
 derivative of the Airy function, namely
\begin{equation}
\label{smallomegaintegral}
I(\omega)\Big|_{\s \omega < 2J}=I_2(\omega)=i\frac{2^{4/3}\pi}{J^{1/3}v^{1/3}}Ai'\Big[\Big(\frac{4J}{v^2} \Big)^{1/3}\left( 2J-\omega\right) \Big].
\end{equation}
The behavior of $I(\omega)$ near $\omega=2J$ is illustrated in
Fig. {\bf \ref{Airyplot}}. For $\omega <2J$ it falls off exponentially
as \\
$\exp\Big[-\frac{2^{7/2}|\nu|}{3}(1-\omega/2J)^{3/2}\Big]$
when $2J-\omega$ exceeds $J/|\nu|^{2/3}$, while for $\omega >2J$ it oscillates and reduces to the asymptote Eq. (\ref{bigomegaIntegral}) after the first maximum. It follows from the plot that, numerically, the small-$\omega$ tail is relatively slim. Still, we will keep it, since it captures $Q_{\s LZ}$ for long
correlation times of the noise.  For arbitrary correlation time, it is sufficient to use the asymptote Eq. (\ref{bigomegaIntegral}) for $\omega>2J$ and the
asymptote Eq. (\ref{smallomegaintegral}) for $\omega <2J$.
Then the expression Eq. (\ref{Fourierprobability}) for the average survival
probability  takes the form
\begin{equation}
\label{QLZintegral}
\langle Q_{\s LZ}\rangle =(\delta J)^2\Bigg[\int\limits_{0}^{2J} d\omega \tilde{K}(\omega) |I_2(\omega)|^2 + \int\limits_{2J}^{\infty} d\omega \tilde{K}(\omega) |I_1(\omega)|^2\Bigg],
\end{equation} where we have replaced $\sinh(\pi|\nu|)$ by $\exp(\pi|\nu|)/2$, since $|\nu|$ is
big.  Eq. (\ref{QLZintegral}) is our main result. While the dependence of $Q_{\s LZ}$ on the
on the noise magnitude is obvious, the dependence on the noise correlation time, predicted by
 Eq. (\ref{QLZintegral}) is nontrivial.
We analyze this dependence in the next section.

 \begin{figure}
 \includegraphics[scale=0.16]{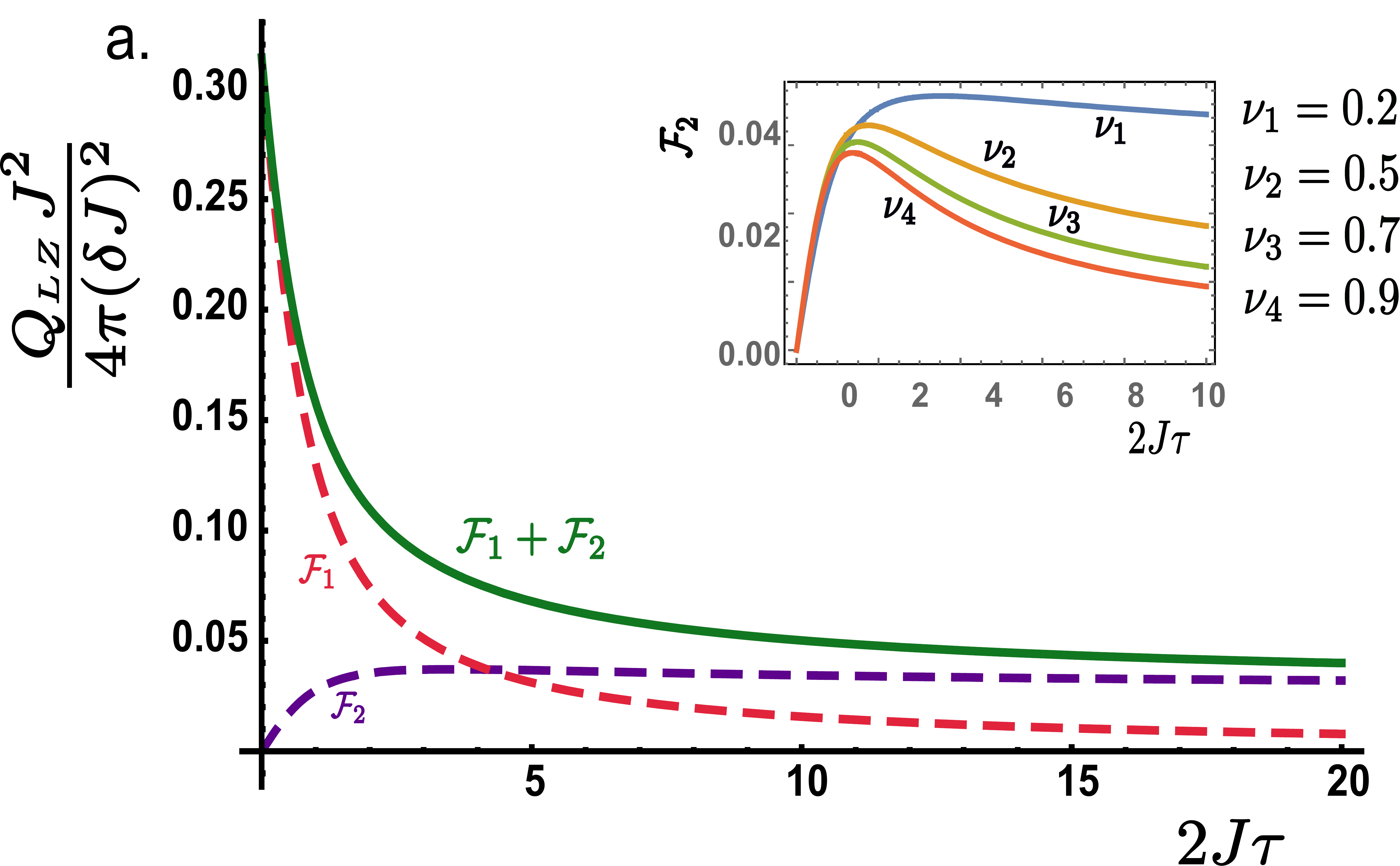}

 \includegraphics[scale=0.16]{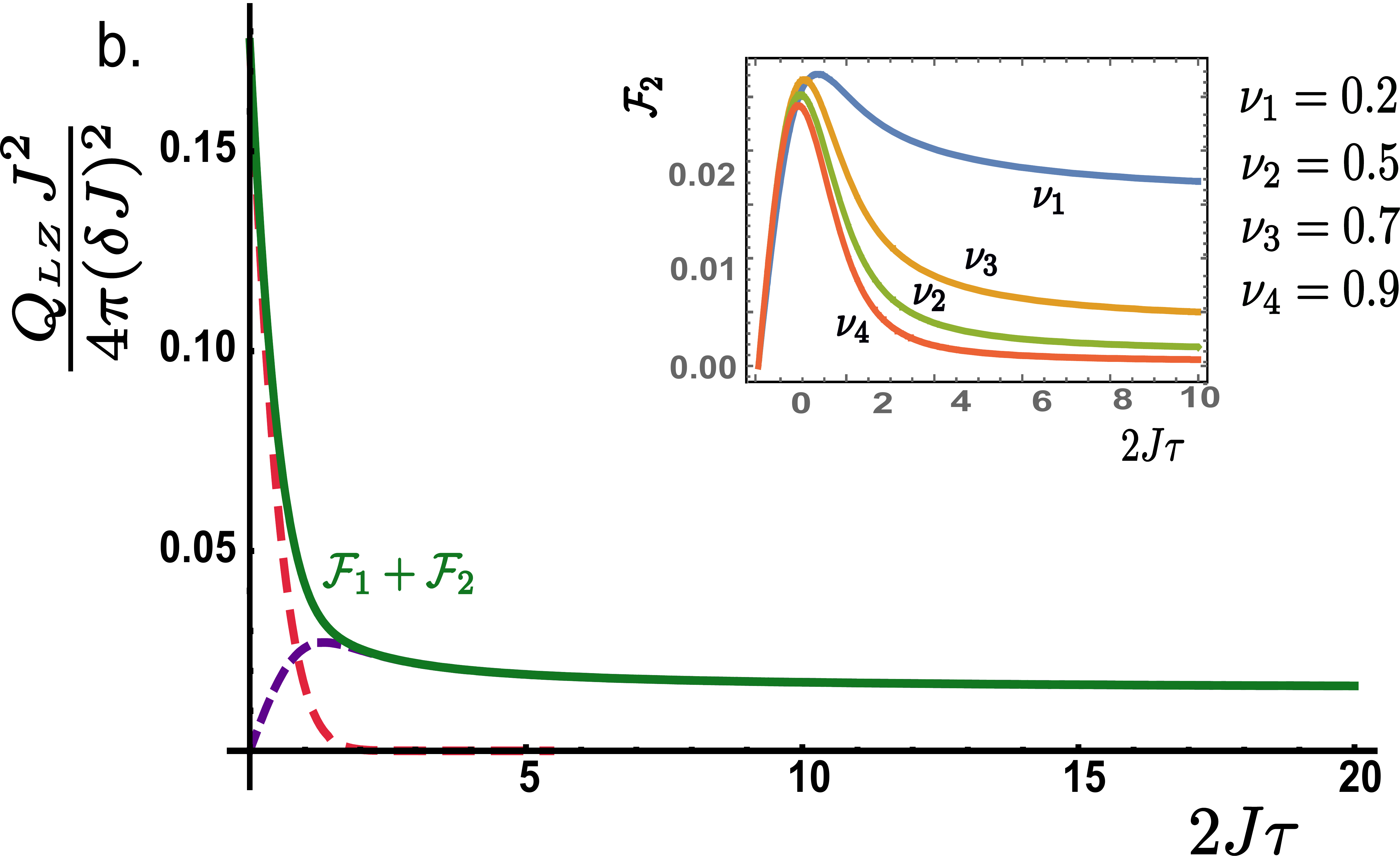}
 \caption{(Color online) Survival probability, $Q_{\s LZ}$, is plotted from Eqs. (\ref{F1+F2})-(\ref{F2})    versus the dimensionless noise correlation time for the telegraph noise (a) and for the gaussian noise (b). The contribution from the absorption of the high-frequency ``noise quanta"  ($\omega>2J$) are shown with red dashed lines (${\cal F}_1$), while the contributions from the absorption with $\omega <2J$
 are shown with purple dashed lines (${\cal F}_2$). The insets illustrate the evolution of the low-frequency contribution with increasing $\nu=J^2/v$. The smaller is the gap, the stronger is the absorption of the sub-gap quanta.  Characteristic correlation time being $\tau \sim 1/J$
 which is much shorter than the time $J/v$ of the LZ transition time indicates that it is ``fast" noise that suppresses the adiabaticity.}
 \label{QLZ_final}
 \end{figure}

\section{Dependence of $\langle Q_{\s LZ} \rangle$ on the noise correlation time  }

If the correlation time of the noise is $\tau$, then $\frac{1}{\tau}{\tilde K}(\omega)$ is a dimensionless
function of the argument $\omega\tau$. Since the frequency scale of both $I_1(\omega)$ and $I_2(\omega)$ is
the gap $2J$, the two contributions to $Q_{\s LZ}$ are the dimensionless functions of the argument $2J\tau$.
Correspondingly, we rewrite Eq. (\ref {QLZintegral}) in the form
\begin{equation}
\label{F1+F2}
\langle Q_{\s LZ} \rangle =   4 \pi \frac{(\delta J)^2}{J^2} \Big[{\cal F}_1(2 J \tau) + {\cal F}_2(2 J \tau) \Big],
\end{equation}
where the functions ${\cal F}_1$ and ${\cal F}_2$ are defined as
 \begin{equation}
 \label{F1}
 {\cal F}_1= |\nu| \int\limits_{2J}^{\infty} d\omega  \tilde{K}(\omega)\Bigl(1-\frac{4J^2}{\omega^2}\Bigr)^{1/2},
 \end{equation}

 \begin{equation}
 \label{F2}
 {\cal F}_2=   2^{2/3} \pi|\nu|^{2/3}\int\limits_{0}^{2J} d\omega
 \tilde{K}(\omega)Ai'^2 \Bigg[    2^{5/3}|\nu|^{1/3}\left( 1-\frac{\omega}{2J}\right)\Bigg]  .
\end{equation}
The first and the second terms describe the absorption of ``above-gap" and ``below-gap" noise quanta, respectively. Note, that the integrand in Eq. (\ref{F1}) does not contain the parameter $\nu$. In Fig. \ref{QLZ_final}(a),(b) we plotted
$Q_{\s LZ}$ for the telegraph noise with ${\tilde K}(\omega)= \frac{\tau}{1+\omega^2\tau^2}$ and for the gaussian noise
with ${\tilde K}(\omega)=\tau \exp(-\omega^2\tau^2)$, respectively.
The contributions ${\cal F}_1$ can be evaluated analytically for both cases.
Namely, for the telegraph noise the calculation yields
\begin{equation}
\label{telegraphF1}
{\cal F}_1(2 J \tau)= \frac{\pi |\nu|}{2}\Biggl(\frac{1}{2J\tau+\left(4J^2\tau^2+1\right)^{1/2}}\Biggr),
\end{equation}
while for gaussian noise the result reads
\begin{equation}
\label{gaussianF1}
{\cal F}_1(2 J \tau)= |\nu|\Bigg[\frac{\pi^{1/2}}{2}\exp{\left(-4J^2\tau^2\right)}
-\pi J\tau~\text{Erfc}(2J\tau)\Bigg],
\end{equation}
where $\text{Erfc}(x)$ is the error function.
The contributions ${\cal F}_1$ dominate $Q_{\s LZ}$ in the small-$\tau$ domain, which corresponds to the fast
noise. In fact, the contribution ${\cal F}_2$  turns to zero for $J\tau \ll 1$. The behavior of the contributions
  ${\cal F}_1$ at small $\tau$ is ${\cal F}_1(2 J \tau)\approx \frac{\pi |\nu|}{2}\left(1-2J\tau \right)$ for the
  telegraph noise and ${\cal F}_2(2 J \tau) \approx |\nu|\bigl(\frac{\pi^{1/2}}{2}-2J\tau \bigr)$ for the gaussian noise.
  The slopes are related as $2/\pi^{1/2}$, i.e. they are close. The fact that for short correlations times the prefactor
  in $Q_{\s LZ}$ is proportional to $\frac{(\delta J)^2}{J^2}|\nu|$ reflects a simple physics that the absorption
  of the high-frequency noise quanta does not depend on $J$. Indeed, $J$ drops out from the  combination $|\nu|/J^2$.

The difference between the two noise realizations manifests itself in the contributions  ${\cal F}_2$. It is
seen from Fig. \ref{QLZ_final} that for the telegraph noise, this contribution falls off with $\tau$ much slower than
for the gaussian noise. In fact, the slow decay of ${\cal F}_2$ can be estimated qualitatively\cite{Luo2017}. Indeed, subsequent jumps of the gap width with magnitude $(\delta J)$ take place at time moments, $t$, separated by $\tau$. A jump results in the absorption only if $t \lesssim J/v$, since $J/v$ is the LZ transition time. The probability that $t \lesssim J/v$ is
$\sim \frac{J}{v\tau}$.  This suggests that ${\cal F}_2$ contribution falls off as $1/J\tau$.
A nontrivial feature of the ${\cal F}_2$ contribution is that it passes through a maximum at $2J\tau \approx 1$.

%
%
\section{Longitudinal noise}

Throughout the paper we assumed that the noise is transverse, i.e. it is described by
the Hamiltonian $\delta J(t)\hat{\sigma}_x$. In this section we briefly outline the
changes to be made in the result Eq. (\ref{F1+F2}) if the noise is longitudinal with
the Hamiltonian $\delta \varepsilon (t)\hat{\sigma}_z$.  The steps of the perturbative
derivation of $\langle|c_2(\infty)|^2\rangle$ leading to Eq. (\ref{bigeq1}) for the longitudinal noise are completely similar to the transverse noise. Naturally, $(\delta\varepsilon)^2$ instead of $(\delta J)^2$ appears in the prefactor. In the
integrand, the combination
$\Big[a_{\s \downarrow}^{\s (1)}(t_1)\Big]^2-\Big[a_{\s \uparrow}^{\s (1)}(t_1)\Big]^2$ gets replaced by $2\Big[a_{\s \downarrow}^{\s (1)}(t_1)a_{\s \uparrow}^{\s (1)}(t_1)\Big]$.
The absolute value of the former combination has a meaning of  $|S_z(t)|$, which is the
 absolute value of the polarization. Correspondingly, the absolute value of the product
 $2\Big[a_{\s \downarrow}^{\s (1)}(t_1)a_{\s \uparrow}^{\s (1)}(t_1)\Big]$ corresponds to
 $|S_x(t)|$. For $|\nu|\gg 1$, this quantity is calculated in the Appendix.
Then the modification of Eq. (\ref{FourierIntegral}) amounts to the replacement of $vt/2$ by $J$ in the numerator of the integrand. As a result,
for $\omega>2J$ the result Eq. (\ref{bigomegaIntegral}) gets modified as
 \begin{equation}
 \label{modifiedI1}
 I_1(\omega)\rightarrow i\frac{2^{5/2}\pi^{1/2}J}{v\left(\omega t_{\omega}\right)^{1/2}}\sin\Big[\omega t_{\omega} -2\Phi(t_{\omega})+\frac{\pi}{4} \Big].
 \end{equation}
Due to this modification, the integral ${\cal F}_1$ in the expression
for the survival probability assumes the form  

\begin{equation}
 \label{modifiedF1}
 {\cal F}_1\rightarrow 4 J^2|\nu|  \int\limits_{2J}^{\infty} d\omega  \frac{\tilde{K}(\omega)}{\omega ^2}\Bigl(1-\frac{4J^2}{\omega^2}\Bigr)^{-1/2}.
 \end{equation}
For the telegraph noise, the evaluation of this integral yields
 
 \begin{equation}
{\cal F}_1(2 J \tau)\rightarrow \frac{\pi |\nu|}{2}\Biggl[\frac{2J\tau}{\left(4J^2\tau^2+1\right)^{1/2}}\Biggr]\Biggl(\frac{1}{2J\tau+\left(4J^2\tau^2+1\right)^{1/2}}\Biggr),
\end{equation}
We see that the result differs from the corresponding expression Eq. (\ref{telegraphF1}) only by an additional factor in the square brackets.
We conclude that the effect of longitudinal noise on $Q_{\s LZ}$
is suppressed, compared to the transverse  noise, in the limit $J\tau \ll 1$, i.e. in the limit of the fast noise.


\section{Discussion}

In this section we compare our results with the results of previous studies\cite{Nalbach2009+, Nalbach2014+, Nalbach2015+, Wubs2006++, Pokrovsky2007++, Wubs2007++,Ao1991} of the effect of noise on the LZ transition.

(i). We calculated the survival probability for arbitrary noise correlation time assuming that the noise is weak,
so that the bare survival probability is exponentially small. This domain of parameters corresponds to ``high-fidelity"
qubit and is most appealing for applications.
In earlier analytical calculations
Refs. \onlinecite{Wubs2006++, Pokrovsky2007++, Wubs2007++} the noise intensity was not assumed to be weak, but the
the noise was assumed to be fast. Both longitudinal and transverse noise were treated on the same footing.
The authors adopted a standard model of a bosonic bath consisting of harmonic oscillators. For the case of
transverse noise (affecting only $J$) considered in the present paper the results of Refs. \onlinecite{Wubs2006++, Pokrovsky2007++, Wubs2007++} can be summarized as follows. In the presence of noise $Q_{\s LZ}=\exp\Big[-2\pi(J^2+(\delta J)^2)/v \Big]$, which suggests that the noise suppresses the survival probability in contrast to what we find.
This conclusion was questioned in subsequent  detailed  numerical studies.\cite{Nalbach2009+, Nalbach2014+, Nalbach2015+}
The results of Refs. \onlinecite{Nalbach2009+}, \onlinecite{Nalbach2014+}, and  \onlinecite{Nalbach2015+}
demonstrate that the Landau-Zener probability decreases with temperature, i.e. with noise magnitude, for all values
of the bare LZ probability (all values of parameter $|\nu|$).
An interesting observation made in these papers is that  $Q_{\s LZ}$, modified by noise, is a non-monotonic function
of $\nu$.

(ii). Technically, our calculation is most close to the paper by Ao and Rammer Ref. \onlinecite{Ao1991}.
In our notations and, within a numerical factor, their result reads,
$Q_{\s LZ}=(\delta J)^2\left(J/v\right)\tilde{K}(2J)n(2J)$, where $n(\omega)$ is the Bose distribution.
The above  expression suggests that the noise-induced survival probability is dominated exclusively by the noise ``quanta"
with frequency $\omega=2J$. This conclusion seems unphysical and contradicts  our result Eq. (\ref{QLZintegral}), according to which all
frequencies with $\omega > 2J$ contribute to  $Q_{\s LZ}$.
On the quantitative level, the difference can be traced to the use of the asymptotes of the parabolic cylinder
functions in Ref. \onlinecite{Ao1991}.

(iii). Note finally, that for very strong noise $\delta J \gg J$
the LZ transition can be viewed as simply noise-driven. This
limit was studied in a pioneering paper Ref. \onlinecite{Kayanuma1985}.
In particular, for fast noise, with frequency much bigger than $v/\delta J$,
the survival probability is given by $Q_{\s LZ}=\frac{1}{2}\Big[1+\exp\left(-4\pi(\delta  J)^2/v\right) \Big]$.

(iv). Throughout the paper we assumed that $\nu$ is big, i.e. the bare survival probability is small.
It is interesting to note that, in the opposite limit of small enough $\nu$, the dependence of survival probability
on the noise magnitude can be {\em non-monotonic}. Below we illustrate this observation
analytically assuming that the noise is slow.

It is known\cite{Kayanuma1985} that in the limit of infinite $\tau_{\s c}$, the average probability of the transition should be calculated by averaging this probability of transition at a given $J$ over the distribution of $J$.

\begin{figure}
\includegraphics[scale=0.22]{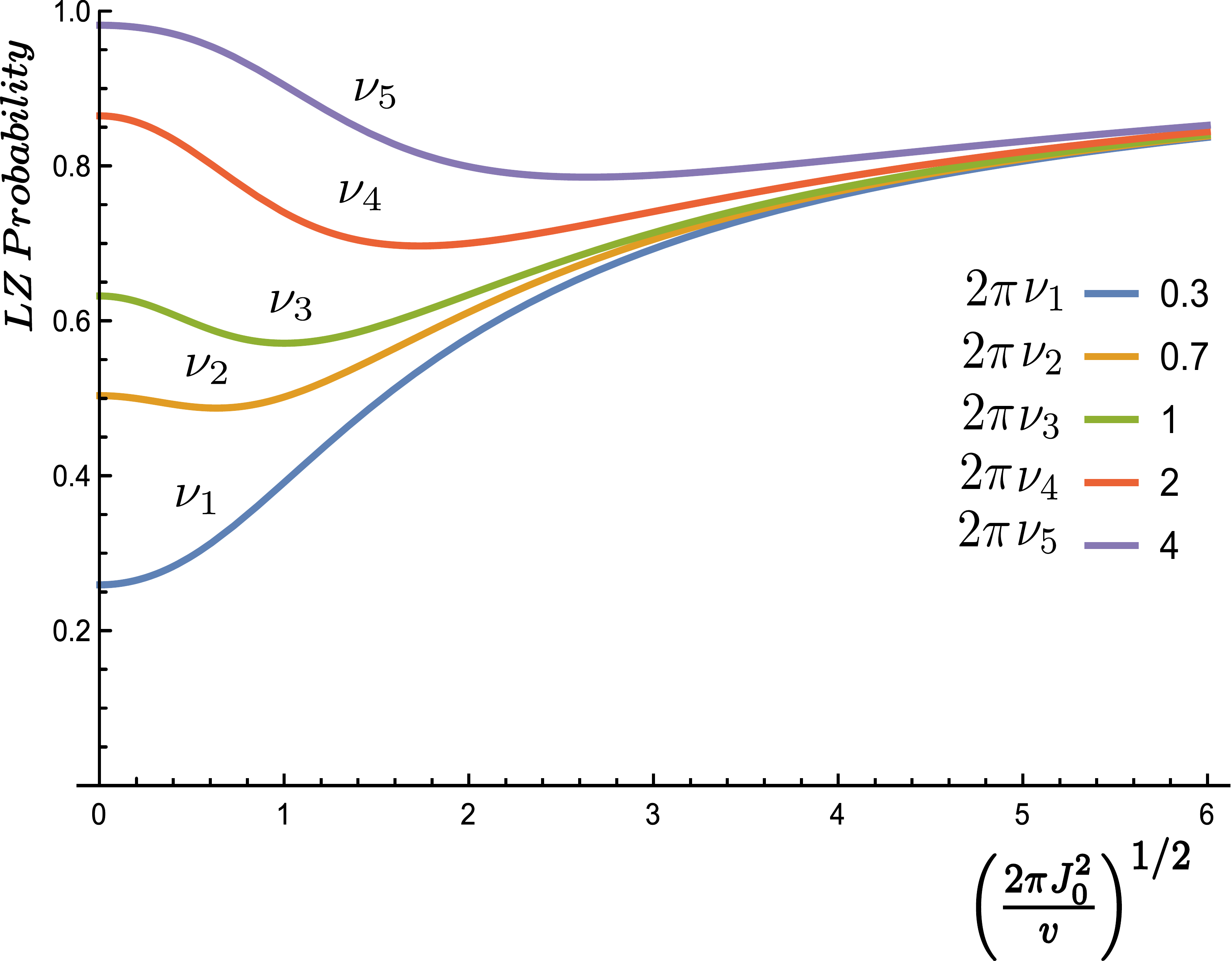}
\caption{(Color online) Landau-Zener probability in the presence of a slow noise is plotted from Eq. (\ref{WithSlowNoise})
as a function of the dimensionless noise magnitude
for different values of parameter $\nu$, which quantifies the bare survival probability $Q_{\s LZ}=\exp(-2\pi|\nu|)$.
For $2\pi |\nu| < 1/2$ the curves grow monotonically, which suggests that noise enhances
the adiabaticity. For $2\pi |\nu| >1/2$, i.e. when the transition is adiabatic in the absence
of noise, the curves exhibit minima, suggesting that a weak noise suppresses the adiabaticity,
while the strong noise, with magnitude exceeding the gap, enhances it. }
\label{LZ1}
\end{figure}

For slow noise with correlation time much longer than $J/v$, the survival probability
is given by
\begin{equation}
\langle Q_{LZ}\rangle =\int\limits_{-\infty}^{\infty}d\delta J~ P\left(\delta J\right)\exp\left[-2\pi|\nu|\left(1+\frac{\delta J}{J}\right)^2 \right].
\end{equation}
For gaussian $P\left(\delta J\right)=\frac{1}{\pi^{1/2} J_0}\exp\big[-\left(\frac{\delta J}{J_0}\right)\big]^2$ the integration yields

\begin{equation}
\label{WithSlowNoise}
\langle Q_{LZ} \rangle =\frac{1}{\big[1+\frac{2\pi}{v}J_0^2\big]^{1/2}}\exp\Big[-\frac{2\pi|\nu|}{1+\frac{2\pi}{v}J_0^2} \Big].
\end{equation}
Note that, for $|\nu| <1/4\pi$, the survival probability is suppressed by noise while for $|\nu| >1/4\pi$
it is   enhanced by noise. This behavior is illustrated in Fig. \ref{LZ1}.
Fig. \ref{LZ1} suggests the following nontrivial effect of low-frequency environment\cite{Ziman2011+} on the LZ transition.
As the coupling to environment, parametrized by $J_0$, increases, the initially adiabatic transition
becomes first less adiabatic, and then, more adiabatic.

(v). The noise spectrum, ${\tilde K}(\omega)$, depends on the concrete
realization of the environment. In theoretical papers, see e.g. Refs. \onlinecite{Nalbach2009+}, \onlinecite{Nalbach2014+}, \onlinecite{Nalbach2015+}, the environment is usually modeled by a set of
harmonic oscillators with the frequency distribution 
$g(\omega)\propto \omega\exp(-\omega/\omega_c)$ (Ohmic environment).
Then ${\tilde K}(\omega)$ is proportional to $g(\omega)\coth(\omega/2T)$,
where $T$ is temperature.

\appendix
\section{Time evolution of the level population in the limit of small survival probability}

In general, the level populations, $P_{\downarrow}(t)$ and  $P_{\uparrow}(t)$
exhibit strong oscillations in the domain $t\sim J/v$,
where the LZ transition takes place. These oscillations originate from the interference of
the terms $\propto \exp (i\Phi(t))$ and  $\propto \exp(-i\Phi(t))$, Eq. (\ref{semiclassicalphase}).
The reason why we were able to find the noise-dependent correction analytically is that, for
small bare survival probability, these oscillations are suppressed.
We established this fact upon analysis of the asymptotes of the parabolic cylinder
functions in the domain $t\sim J/v$.
It is instructive to trace how the result Eq. (\ref{numeratorbigeq})
\begin{equation}
\label{A1}
P_{\downarrow}(t)-P_{\uparrow}(t)=\frac{\frac{vt}{2}}{(J^2+\frac{v^2t^2}{4})^{1/2}}
\end{equation}
emerges  from the alternative description based on the  spin dynamics.
In the literature, the effect of noise on the LZ transition is studied within
this description.

The difference $P_{\downarrow}(t)-P_{\uparrow}(t)=S_z(t)$ can be viewed as spin
polarization, while the system Eq. (\ref{system}) describes the
evolution of the $\uparrow$ and $\downarrow$ spin amplitudes
in the effective magnetic field, $\bm B$,  with components $B_z(t)=\frac{vt}{2}$ and $B_x=J$.
Three equations of motion
for the spin projections following from $\frac{d{\bm S}}{dt}={\bm B}\times{\bm S}$ can be reduced to a single integral-differential
equation for $S_z(t)$
\begin{equation}
\label{A2}
\frac{\partial S_z(t)}{\partial t}=-\int\limits_{-\infty}^t dt' \cos\left(\int\limits_{t'}^t dt'' B_z(t'')\right)
B_x(t)B_x(t')S_z(t').
\end{equation}
The crucial simplification, which allows to solve this equation in the limit $|\nu|\gg 1$ is that, for relevant times $t\sim J/v$,
the argument of cosine $\int\limits_{t'}^t dt'' B_z(t'')=\frac{v}{4}(t^2-t'^2)$ is big. For $B_x(t)=B_x(t')=J$, Eq. (\ref{A2}) takes the form
\begin{equation}
\label{A3}
\frac{\partial S_z(t)}{\partial t}=-J^2\int\limits_{-\infty}^t dt' \cos\left[\frac{v}{4}(t^2-t'^2)\right]
S_z(t').
\end{equation}
Strong oscillations of cosine suggest that the major contribution to the integral comes from $(t-t')\ll t$. To make use of this condition, we perform the integration by parts in the right-hand side
\begin{equation}
\label{A4}
\frac{\partial S_z(t)}{\partial t}=-\frac{2J^2}{v}\int\limits_{-\infty}^t dt' \sin\left[\frac{v}{4}(t-t')(t+t')\right]
\frac{\partial \left(\frac{S_z(t')}{t'}\right)}{\partial t'}.
\end{equation}
Next, we set $t+t'=2t$ in the argument of sine and set $t=t'$ in the derivative. This yields
\begin{equation}
\label{A5}
\frac{\partial S_z(t)}{\partial t}=-\frac{2J^2}{v}\frac{\partial \left(\frac{S_z(t)}{t}\right)}{\partial t}\int\limits_{-\infty}^t dt' \sin\left[\frac{v}{2}(t-t')t\right].
\end{equation}
Now the integration over $t'$
can be carried out leading to
\begin{equation}
\label{A6}
\frac{\partial S_z(t)}{\partial t}=-\frac{4J^2}{v^2 t}\frac{\partial \left(\frac{S_z(t)}{t}\right)}{\partial t}=-\frac{4J^2}{v^2 t}\Bigg[\frac{1}{t}\frac{\partial S_z(t)}{\partial t}-\frac{S_z(t)}{t^2} \Bigg].
\end{equation}
The first order differential equation Eq. (\ref{A6}) can be easily solved. With initial condition $S_z(-\infty)=-1$, the result reads
\begin{equation}
\label{A7}
S_z(t)=\frac{\frac{vt}{2}}{\left(J^2+\frac{v^2t^2}{4}\right)^{1/2}}=\frac{B_z}{\left(B_x^2+B_z^2\right)^{1/2}},
\end{equation}
i.e. the polarization is equal to cosine of the angle between magnetic field and the $z$-axis. Using Eq. (\ref{A7}), the projection $S_y(t)$  can be calculated from the equation $\frac{dS_z}{dt}=B_xS_y$ and turns out to be
\begin{equation}
\label{A8}
S_y(t)=\frac{J \frac{v}{2}}{\left(J^2+\frac{v^2t^2}{4}\right)^{3/2}}=\frac{B_x \frac{\partial B_z}{\partial t}}{\left(B_x^2+B_z^2\right)^{3/2}}.
\end{equation}
Subsequently, the projection $S_x(t)$ calculated from  $\frac{dS_x}{dt}=-B_zS_y$ acquires the form
\begin{equation}
\label{A9}
S_x(t)=\frac{J}{\left(J^2+\frac{v^2t^2}{4}\right)^{1/2}}=\frac{B_x}{\left(B_x^2+B_z^2\right)^{1/2}}.
\end{equation}
From the expressions Eqs. (\ref{A7})-(\ref{A9}), we can estimate the accuracy of the approximations made.
These expressions are valid if $S_y \ll 1$. Indeed, it follows from (\ref{A7}), (\ref{A9}) that $S_z^2+S_x^2=1$. On the other hand, it follows
from Eq. (\ref{A8}) that the maximal value of $S_y$ is $\frac{v}{J^2}=\nu^{-1}\ll 1$. Thus, the results Eqs. (\ref{A7})-(\ref{A9})
are valid with accuracy $\nu^{-1}$. Uncertainty $\sim \nu^{-1}$ is much bigger than the inaccuracy of the result $S_z(\infty)=1$, which follows
from Eq. (\ref{A7}). Inaccuracy of this result is $\exp(-2\pi \nu)$, i.e.
it is exponentially small.

Numerical results for the spin projections in the limit $\nu \gg 1$ are
presented in Ref. \onlinecite{Ziman2011+}.  They seem to be in good agreement with analytical expressions Eqs. (\ref{A7})-(\ref{A9}).

\centerline{\bf Acknowledgements}
Illuminating discussion with V. L. Pokrovsky is gratefully acknowkedged.
The work was supported by the Department of
Energy, Office of Basic Energy Sciences, Grant No. DE-
FG02-06ER46313.

\end{document}